\documentstyle[prb,preprint,eqsecnum,aps]{revtex}
\tighten
\begin{document}
\draft
\title{Topology of spacetime and quantization of particle dynamics}
\author{W\l odzimierz Piechocki\thanks{Electronic mail: 
piech@fuw.edu.pl } \\}\address{So\l tan Institute for Nuclear 
Studies, Ho\.{z}a 69, 00-681 Warszawa, Poland\\}
\date{\today}
\maketitle

\begin{abstract}

We compare classical and quantum dynamics of a particle in the de Sitter 
spacetimes with different topologies to show that the result of quantization 
strongly depends on global properties of a classical system. We present 
essentially self-adjoint representations of the algebra of observables for 
each system. Quantization based on global properties of a classical system 
accounts properly its symmetries. 
\end{abstract}
\narrowtext

\section{INTRODUCTION}

Recently\cite{GW} we have found that classical and quantum dynamics 
of a free particle in a curved spacetime seems to be  sensitive 
to the topology of spacetime. The existence of such a dependence is 
interesting not only in itself but may be important for quantum gravity 
and quantum cosmology. The aim of the present paper is examination of  
this dependence in more details to understand its essence. 

For simplicity we restrict ourselves to  the dynamics of a free particle 
in two-dimensional spacetimes. In Sec. IV  we make some comments concerning 
more general cases.
The considered spacetimes,  $V_p$ and $V_h$, are of de Sitter's type. 
They are defined to be
\begin{equation}
V_p = (R^1\times R^1,~\hat{g})~~~~~~\text{and}~~~~~~~~
V_h = (R^1\times S^1,~\hat{g}).
\end{equation}
In both cases  the metric $g_{\mu\nu} :=(\hat{g})_{\mu\nu} ~~(\mu ,\nu= 0,1)$ 
is defined by the line-element
\begin{equation}
ds^2 = dt^2 - exp(2t/r)~dx^2,
\end{equation}
where $r$ is a real positive constant.

\noindent
It is clear that (1.1) presents all possible topologies 
of de Sitter's type spacetimes in two dimensions.
$V_p$ is a plane with global $(t,x)\in R^2$ 
coordinates. $V_h$ is defined to be a one-sheet 
hyperboloid embedded in 3d Minkowski space. There exists an isometric 
immersion map\cite{ES} of $V_p$ into $V_h$
\begin{equation}
 V_{p} \ni (t,x)\longrightarrow (y^0,y^1,y^2) \in V_h,
 \end{equation}
where
$$ y^0:=r\sinh (t/r) +\frac{x^2}{2r}~\exp(t/r),~~~
y^1:=-r\cosh (t/r)+\frac{x^2}{2r}~\exp(t/r),~~~
 y^2:= -x \exp(t/r),$$
and where 
\begin{equation}
 (y^2)^2+(y^1)^2-(y^0)^2=r^2.
\end{equation}
Eq. (1.3) defines a global map of $V_p$ onto a simply connected 
noncompact region of $V_h$. One can check that the induced metric on 
$V_h$ coincides with the metric defined by (1.2).

It is known\cite{HE} that $V_p$ is geodesically incomplete. However, all 
incomplete geodesics in $V_p$ can be extended to complete ones in $V_h$, 
i.e. $V_p$ has removable type singularities. $V_p$ and $V_h$ are 
the simplest examples of spacetimes with noncompact and compact spaces, 
respectively, and with constant curvatures $R= -2 r^{-2}$.

\section{Classical dynamics}

The action integral, $S$, describing a free relativistic particle of 
mass $m$  in gravitational field  $g_{\mu \nu}~~ (\mu,\nu =0,1)$ is 
proportional to the length of a particle world-line and is given by
\begin{equation}
S=\int_{\tau_1}^{\tau_2}~L(\tau)~d\tau,~~~~~L(\tau):=-m\sqrt{g_{\mu\nu}
(x^0(\tau),x^1(\tau))\;\dot{x}^\mu (\tau)\dot{x}^\nu (\tau)},
\end{equation}
where $\tau$ is an evolution parameter, $x^\mu$ are 
spacetime coordinates and $\dot{x}^\mu := dx^\mu/d\tau $. It is assumed 
that $ \dot{x}^0 >0 $, i.e., $x^0$ has interpretation of time monotonically 
increasing with $\tau$. 

The Lagrangian  (2.1) is invariant under the reparametrization
$ \tau\rightarrow f(\tau)$.  
This gauge symmetry leads to the constraint 
\begin{equation}
 G:= g^{\mu\nu}p_\mu p_\nu -m^2=0,
\end{equation} 
where $g^{\mu\nu}$ is an inverse of $g_{\mu\nu}$ and 
$p_\mu := \partial L/\partial\dot{x}^\mu$ are canonical momenta. 

Since we assume that a free particle does not modify the geometry of 
spacetime, the local symmetry of the system is defined by the set of all Killing 
vectors of  spacetime (which is also the symmetry of the 
Lagrangian $L$). The corresponding dynamical integrals have the form\cite{VA}
\begin{equation} 
D=p_\mu X^\mu,~~~~~\mu= 0,1 , 
\end{equation}
where $X^\mu$ is a Killing vector field.

\noindent 
The physical phase-space $\Gamma$ is defined to be the space of all particle 
trajectories consistent with the dynamics of a particle and with the 
constraint (2.2).

\subsection{Dynamics on hyperboloid}

It is known that the hyperboloid (1.4) is invariant under the proper 
Lorentz transformations, i.e. $SO(1,2)$ is the symmetry group of $ V_h $
 system. The infinitesimal transformations of $SO(1,2)$ group  
with parametrization of  the hyperboloid (1.4) in the form
\begin{equation}
y^0=-\frac{r\cos \rho /r}{\sin \rho /r},~~~~y^1=\frac{r\cos \theta /r}
{\sin \rho /r},~~~~y^2=\frac{r\sin \theta /r}{\sin \rho /r},
~~~~\rho \in ~]0,\pi r[,~~~\theta \in [0,2\pi r[,
\end{equation} 

\noindent
read
\begin{eqnarray}
(\rho,~\theta)\longrightarrow(\rho,~\theta+a_0 r),~~~~~~~~~~~~~~~~~~~~
\nonumber \\
(\rho,~\theta)\longrightarrow(\rho-a_1 r \sin \rho /r~\sin 
\theta /r,~\theta+a_1 r\cos \rho /r~\cos \theta /r),\nonumber \\
(\rho,~\theta)\longrightarrow(\rho+a_2 r\sin \rho /r~\cos 
\theta /r,~\theta+a_2 r\cos \rho /r~\sin \theta /r),
\end{eqnarray}
where $(a_0,a_1, a_2) \in R^3 $ are parameters and belong to the neighborhood 
of $(0,0,0) \in R^3$.

\noindent
The corresponding dynamical integrals (2.3) are
\begin{eqnarray}
J_0=p_{\theta}~r,~~~~~J_1=-p_\rho ~r\sin \rho /r~\sin 
\theta /r + p_\theta ~r\cos \rho /r~\cos \theta /r, \nonumber \\
J_2= p_\rho ~r\sin \rho /r~\cos \theta /r 
+ p_\theta ~r\cos \rho /r~\sin \theta /r,~~~~~~~~~~~~
\end{eqnarray}
where $p_\theta :=\partial L/\partial\dot{\theta},~p_\rho :=
\partial L/\partial\dot{\rho}$ are canonical momenta.

One can check that the dynamical integrals (2.6) satisfy the commutation 
relations of $sl(2,R)$ algebra
\begin{equation}
\{ J_a , J_b \} =\varepsilon_{abc}\eta^{cd} J_d ,
\end{equation}
where $\varepsilon_{abc}$
is the anti-symmetric tensor with $\varepsilon_{012}=1$ and $ \eta^{cd} $ 
is the Minkowski metric tensor.

The constraint (2.2), in terms of dynamical integrals, reads
\begin{equation}
J_0^2 - J_1^2 - J_2^2 = -\kappa^2,
\end{equation}
where $\kappa =mr$.

\noindent
Eqs. (1.4), (2.4) and (2.6) define particle trajectories
\begin{equation}
J_a y^a =0,~~~~~J_2 y^1 - J_1 y^2 =r^2 p_\rho ,
\end{equation}
where $p_\rho < 0 $ since we consider time-like trajectories $(|\dot{\rho}|>
|\dot{\theta}|,~\dot{\rho}>0 )$.

Each point $(J_0, J_1, J_2)$ of (2.8) defines uniquely a particle 
trajectory (2.9) on (1.4) admissible by the dynamics and consistent with 
the constraint (2.2). 
Thus, the one-sheet hyperboloid (2.8) defines the physical phase-space 
$\Gamma_h$ and $SO(1,2)$ is  the symmetry group of $\Gamma_h$.  
Since $sl(2,R)$ is the Lie algebra isomorphic to the Lie algebra of $SO(1,2)$ 
group, we have a clear relationship between local and global symmetries of 
the  classical $V_h$ system.

\subsection{Dynamics on plane}

The Lagrangian (2.1) with the metric tensor defined by (1.2) reads
\begin{equation}
L=-m\sqrt{\dot{t}^2- \dot{x}^2\exp(2t/r)}, 
\end{equation}
where $t:=x^0,~x:=x^1,~\dot{t}=dt/d\tau$ and $\dot{x}=dx/d\tau$.

\noindent
The local symmetry of L is defined by translations
\begin{equation}
(t,~x)\longrightarrow (t,~x+b_0),
\end{equation}
space dilatations with time translations
\begin{equation}
(t,~x)\longrightarrow (t-rb_1,~x+xb_1 ) ,
\end{equation}
and by the transformations
\begin{equation}
(t,~x)\longrightarrow (t-2rxb_2 ,~x+ (x^2 +r^2 e^{-2t/r})b_2),
\end{equation}
where $(b_0,b_1,b_2) \in R^3 $ are parameters and belong to the neighborhood 
of $(0,0,0) \in R^3 $.

\noindent
The Killing vector fields corresponding to the transformations (2.11), 
(2.12) and (2.13)  define, respectively, the dynamical integrals (2.3)
\begin{equation}
 P=p_x,~~~K=-rp_t+xp_x,~~~M=-2rxp_t +(x^2 + r^2 e^{-2t/r}) p_x ,
\end{equation}
where $p_x = \partial L/\partial\dot{x},~ p_t = \partial L/\partial \dot{t}$. 

\noindent
One can verify that the dynamical integrals (2.14) satisfy the commutation 
relations of $sl(2,R)$ algebra
\begin{equation}
\{P,K\}=P,~~~\{K,M\}=M,~~~\{P,M\}=2K.
\end{equation}

\noindent
The mass-shell condition (2.2) takes the form
\begin{equation}
p^2_t -e^{-2t/r}p^2_x =m^2,
\end{equation}
 which, due to (2.14), relates the dynamical integrals   
\begin{equation}
K^2-PM=\kappa ^2,~~~~\mbox {where}~~~~ \kappa =mr.
\end{equation}

Similarly to the case of $V_h$ system, the dynamical integrals (2.14) 
satisfying (2.17) should determine particle trajectories. However, not 
all such trajectories are consistent with particle dynamics:

\noindent
For $P=0$ there are two lines $ K=\pm \kappa$ on the hyperboloid (2.17).
Since by assumption $\dot{t} >0$, we have that $p_t = \partial L/
\partial\dot{t}=-m\dot{t}\:(\dot{t}-\dot{x}\exp(2t/r))^{-1/2} < 0 $. 
According to (2.14) $K-xP = -rp_t,$ thus $K-xP > 0,~$ i.e. $K>0~$ for $P=0$.
Therefore, the line $(P=0,~K=-\kappa)$ is not available for the dynamics. 
The hyperboloid (2.17) without this line defines the physical phase-space 
$\Gamma_p$.
 
Excluding the momenta $p_t$ and $p_x$ from (2.14) we find the  particle 
trajectories (parametrized by $\Gamma_p$)
\begin{equation}
x=M/2K,~~~~~~~~~~~~~~~~~~~\mbox{for}~~~~P=0
\end{equation}
and 
\begin{equation}
xP=K-\sqrt{\kappa^2 +(rP)^2\exp{(-2t/r)}},~~~~~\mbox{for}~~~~P\neq 0,  
\end{equation}
where (2.19) takes into account that $K-x P>0.$

It is clear that $\Gamma_p$ is topologically equivalent to a plane 
$R^1\times R^1$. 
We can parametrize $\Gamma_p$  by the coordinates $(q,p)\in R^1 \times R^1$  
as follows
\begin{equation}
P=p,~~~K=pq-\kappa,~~~M=pq^2 -2\kappa q.
\end{equation}
Eqs. (2.15) and (2.20) give 
\[ \{p,q\}=1,~~~~~\{p,p\}=0=\{q,q\} . \] 

The local symmetry of the system is defined by $sl(2,R)$ algebra. 
However, the $SO(1,2)$ group cannot be the global symmetry of the system 
because $V_p$ is only a subspace of $V_h$ due to the isometric immersion 
map (1.3). In fact, the Killing vector field generated by the transformation 
(2.13) is not complete on $V_p$, whereas the vector fields generated by (2.11) 
and  (2.12) are well defined globally (see App. A).   
Therefore, the global symmetry of the system is  the Lie group with the 
Lie algebra defined by the commutation relation
\begin{equation}
\{P,K\} = P.
\end{equation}
Eq.(2.21) defines a solvable subalgebra of $sl(2,R)$ algebra.

In case of $V_p$ system the Lie algebra corresponding to the global 
symmetry of $V_p$ is different from  $sl(2,R)$ algebra of all 
available Killing vector fields. This breaks the nice relationship between 
local and global symmetries which occurs in $V_h$ case.

\subsection{Observables and phase-spaces coordinates}

The classical observables are  defined to satisfy the following conditions: 

\begin{description}
\item[(i)] algebra of observables corresponds to local symmetry  
          of $V_p$ or $V_h$ system; 
\item[(ii)] observables specify particle trajectories admissible by dynamics  
            ($V_p$ and $V_h$ are integrable systems);
\item[(iii)] observables are gauge invariant, i.e., have vanishing Poisson's 
             brackets with the constraint $G$, Eq.(2.2). 
\end{description}

We apply  the group quantization method\cite{CJI}.  The coordinates 
on $\Gamma_p$ and $\Gamma_h$ are chosen to satisfy the two conditions: 
(a) symplectic structure on $\Gamma_p$ or $\Gamma_h$  has canonical 
form and (b) classical observables are first order polynomials 
in one of the canonical coordinates. Such a choice enables, in the 
quantization procedure, solution of the operator-ordering problem by 
symmetrization and simplifies discussion of self-adjointness of quantum 
operators. 

The analyses we have done so far show that the physical phase-space 
of each system has the same symmetry as the corresponding spacetime. 
Since considered spacetimes are globally isometric at the intersection of
$V_h$ with the range of the map (1.2), the difference between  physical 
phase-spaces results from the difference between spacetime topologies 
of $V_{p}$ and $V_h$ systems. 

\section{QUANTIZATION}

By quantization we mean here finding an essentially self-adjoint 
representation of the algebra  of classical observables on a dense subspace 
of a Hilbert space (considered quantum observables are unbounded operators).
 
It is a starting point for further analysis. In particular, one can examine 
the integration of the algebra representation to the unitary representation 
of the symmetry group of a classical system.

\subsection{Quantum dynamics on hyperboloid}

We choose $J_0,J_1$ and $J_2$ as the classical observables. 
One can easily verify that the criteria (i)-(iii) of Sec. IIC are satisfied. 
To meet the conditions (a) and (b) of Sec. IIC we parametrize the hyperboloid 
(2.8) as follows
\begin{equation}
J_0=J,~~~J_1=J\cos\beta - \kappa\sin\beta,~~~J_2=-J\sin\beta 
-\kappa\cos\beta ,
\end{equation}
where $J\in R^1$ and $\beta\in S^1.$

\noindent
One can check that the canonical commutation relations 
\[ \{\beta ,J\}=1,~~~~~\{\beta,\beta\}=0=\{J,J\} \]
lead to (2.7).

Making use of the Schr\"{o}dinger representation for the canonical 
coordinates $\beta$ and $J$ (see App. B) and applying the symmetrization 
prescription to (3.1) we obtain 
\begin{equation}
\hat{J}_0\psi(\beta)=\frac{\hbar}{i}\frac
{d}{d\beta}\psi(\beta) ,
\end{equation}
\begin{equation}
{\hat{J}}_1 \psi(\beta)= \Bigl( \cos \beta 
~\hat{J_0} - ({\kappa} -\frac{i\hbar}{2})\sin \beta\Bigr)\psi(\beta),
\end{equation}
\begin{equation}
\hat{J}_2 \psi(\beta)= \Bigl( -\sin \beta 
~\hat{J_0} - ({\kappa} -\frac{i\hbar}{2})\cos \beta\Bigr)\psi(\beta) ,
\end{equation}
where $\psi \in\Omega \subset L^2[0,2\pi]$.

The subspace $ \Omega$ is defined to be
\begin{equation}
\Omega := \{\psi\in L^2[0,2\pi]~|~\psi\in C^\infty[0,2\pi],~\psi^{(n)}(0)
= \psi^{(n)}(2\pi),~ n=0,1,2...\} .
\end{equation}
It is clear that $\Omega$ is a common invariant dense domain for 
$ \hat{J}_a~~(a=0,1,2)$.

\noindent
One can verify that 
\begin{equation}
[\hat{J}_a,\hat{J}_b]\psi =\frac{\hbar}{i}\widehat{\{J_a,J_b\}}\psi,
~~~~\psi\in\Omega ,
\end{equation}
and that the representation (3.2-3.5) is symmetric on $\Omega$, 
if $\kappa$ is real. We prove in the Appendix C that the representation is 
essentially self-adjoint.

It is interesting to mention that   presented representation of $sl(2,R)$ 
algebra is essentially self-adjoint despite of the fact that representation 
of the canonical commutation relations can be at most symmetric (see App. B).

\subsection{Quantum dynamics on plane}

To compare the dynamics of $V_p$ and $V_h$ systems at the quantum 
levels we apply to the phase-space of $V_p$ system the symplectic 
transformation $(q, p)\rightarrow (\sigma, I)$ defined by
\begin{equation}
q:=-\cot\frac{\sigma}{2},~~~p:=(1-\cos\sigma)(I+\kappa\cot\frac{\sigma}{2}),
\end{equation}
where $0<\sigma<2\pi$ and $I\in R^1$.

\noindent
The dynamical integrals (2.14) rewritten in $(\sigma,I)$ variables lead to 
\[ I_0:=\frac{1}{2}(M+P)=I,\]
\begin{equation}
I_1:=\frac{1}{2}(M-P)=I\cos\sigma -\kappa\sin\sigma,
~~~~~I_2:=K=-I\sin\sigma -\kappa\cos\sigma.
\end{equation}
The commutation relations for $I_a~(a=0,1,2)$ resulting from (2.15) have the 
form (2.7). Comparing (3.8) with (3.1) we see that $I_a$ and $J_a$ have 
the same functional form. However, they are  defined on 
topologically different phase-spaces $\Gamma_p$ and $\Gamma_h$, where

\begin{equation}
\Gamma_p =(0,2\pi)\times R^1,~~~~~\Gamma_h =S^1 \times R^1 .
\end{equation}
The quantization of $V_p$ system can be done by analogy to the $V_h$ case. 
The only difference is that the quantum operators $\hat{I}_a~(a=0,1,2)$ 
corresponding to (3.8) have different domain. One can  check that 
now a dense invariant common domain of $\hat{I}_a$ can be taken to be 
\begin{equation}
\Omega_\alpha:=\{\psi\in L^2[0,2\pi]~|~\psi\in C^\infty [0,2\pi],~\psi^{(n)}
(0)=e^{i\alpha}\psi^{(n)}(2\pi),~n=0,1,2,...\},
\end{equation}
where $0\leq \alpha <2\pi$, i.e. the representation of (3.8) is parametrized by 
a continuous real parameter $\alpha$.

\noindent
The case $\alpha=0$ corresponds to the representation of $sl(2,R)$ algebra 
of $J_a~(a=0,1,2)$ observables.  
In case of $V_h$ system the choice $\alpha=0$ results from the fact that 
 $\beta=0$ and $\beta=2\pi$ for fixed $J$, in parametrization (3.1), label 
the same point of the hyperboloid (2.8). 

\noindent
In case of $V_p$ system the end points 
of the range of $\sigma$ in (3.8) do not coincide. Thus, there is no reason 
to choose any specific value for $\alpha$.
Each choice of $\alpha$ defines an essentially  self-adjoint representation 
of $sl(2,R)$ algebra of $I_a~ (a=0,1,2)$ observables (see, App.C).  
Thus, we have infinitely many unitarily nonequivalent quantum $V_p$ systems 
corresponding to just one $V_p$ classical system. This can be treated at the 
best as a phenomenological model with $\alpha$ as a free parameter, but such 
a quantum theory has no predictability. Also it 
is unclear how to extract from this result the representation of the algebra 
(2.21) useful for finding a unique unitary representation of the symmetry 
group of $V_p$ system. 

Since we are interested in finding a fundamental theory, this approach to 
quantization is not satisfactory. We propose another method which 
consists in replacement of the condition {\bf(i)} of Sec. IIC by the 
following: 

{\bf (i)} algebra of observables corresponds to the symmetry group of 
$V_p$ or $V_h$ system; 

\noindent
This change brings nothing new for $V_h$ case, since the Lie algebra of the 
symmetry group $SO(1,2)$ is isomorphic to $sl(2,R)$ algebra. There are 
substantial changes in $V_p$ case.
The set of observables, due to (2.21,) consists now of only two observables 
$P$ and $K$. It appears that the condition (ii) of Sec. IIC is violated, since 
to specify a particle trajectory, Eqs. (2.18) and (2.19), we also need $M$ 
integral. In fact $M$ can be calculated from Eq.(2.17) because $P,K$ and 
$M$ integrals are not independent.

The observables $P$ and $K$, Eq. (2.20),  are already 
linear in both canonical coordinates $q$ and $p$.
To find the quantum operators corresponding to $P$ and $K$  we 
use the  Schr\"{o}dinger representation for the canonical coordinates $q$ 
and $p$, and apply the symmetrization method. As the result we have 
\begin{equation}
\hat{P}=\frac{\hbar}{i}\frac{d}{dq},~~~~\hat{K}=\frac{\hbar}{i}
q\frac{d}{dq}+ \frac{\hbar}{2i}-\kappa
\end{equation}
where the common invariant dense domain, $\Lambda$, for $\hat{P}$ and 
$\hat{K}$ is defined to be 
\begin{equation}
\Lambda := \{~\psi\in L^2 (R)~|~\psi\in C_0^\infty (R)~\}.
\end{equation}
One can verify that 
\begin{equation}
[\hat{P},\hat{K}]=\frac{\hbar}{i}\hat{P}
\end{equation}
and that both $\hat{P}$ and $\hat{K}$ are symmetric on $\Lambda$, if $\kappa$ 
is real. 
In fact $\hat{P}$ and $\hat{K}$ are essentially self-adjoint on $\Lambda$ 
(see App. D). Eqs. (3.11) and (3.12) define an essentially 
self-adjoint representation of the algebra (2.21).  It can be further examined 
for its integrability to the unitary representation of the symmetry group.

\section{CONCLUSIONS}
Local properties of a given spacetime like metric tensor and Lie algebra 
of the Killing vector fields do not specify the system uniquely because 
systems with  different transformation groups may have isometric Lie 
algebras\cite{LP,BR}.  Also spacetimes with the same local properties 
may have different topologies and such that algebras 
corresponding to the transformation groups  may differ from the algebras 
of all the Killing vector fields. Presented results show that the topology of 
spacetime is a basic characteristic of a classical system. It codes global 
and (indirectly) local symmetries, and also singularities of spacetime.

The results of quantization of $V_p$ and $V_h$ systems are drastically 
different despite of the fact that at the classical level the systems are 
locally identical. The incomplete geodesics of $V_p$ spacetime create 
problems difficult to deal with at the quantum level. 
There are no problems when quantizing the $V_h$ system. There exists in this 
case a straightforward  relationship between local and global symmetries both 
at classical and quantum levels. Quantization seems to favor spacetimes with 
complete geodesics.

Our recent analyze has shown\cite{H4} that one can generalize presented 
results to the de Sitter spacetimes with topologies 
$(R^1 \times R^3, ~\hat{g})$ 
and $(R^1 \times S^3, ~\hat{g})$, since there exists corresponding to (1.3) 
isometric immersion map. It is clear that one can meet similar problems 
when considering a particle dynamics in any spacetime with topology 
admitting removable type singularities.

We expect that quantization of a particle dynamics in spacetimes with 
topologies admitting essential type singularities\cite{JS} may bring some 
new insight into the problem of consolidation of quantum mechanics with 
general relativity.

\section*{Acknowledgments}

I would like to thank George Jorjadze for numerous valuable discussions on 
dynamics of considered systems. I also thank  Prof. S.L. Woronowicz for 
helpful discussions concerning the problem of self-adjointness of 
unbounded operators.   


\appendix
 
\section{GLOBAL TRANSFORMATIONS ON PLANE}

The transformations (2.11), (2.12) and (2.13) lead, respectively, 
to the following infinitesimal generators\cite{NHI} 
\begin{equation}
X_1 = \partial/\partial x,
\end{equation}
\begin{equation}
X_2 = -r\partial/\partial t + x \partial/\partial x,
\end{equation}
\begin{equation}
X_3 = -2rx\partial/\partial t + (x^2 + r^2 \exp(-2t/r))\partial/\partial x .
\end{equation}
The one-parameter group generated by $X_3$ is defined by the solution 
of the Lie equations 
\begin{equation}
\frac{dt}{db_3}=-2rx,
\end{equation}
\begin{equation}
\frac{dx}{db_3}=x^2 +r^2 \exp(-2t/r) ,
\end{equation}
\begin{equation}
t_{\mid b_1 =0= b_2 =b_3} =t_0
\end{equation}
\begin{equation}
x_{\mid b_1 =0= b_2 =b_3} = x_0. 
\end{equation}
(In what follows we use $\epsilon :=b_3$ to simplify notation.)

Acting of $\partial/\partial\epsilon $ on (A5) and making use of (A4) gives
\begin{equation}
\frac{d^2 x}{d\epsilon ^2}-6x\frac{dx}{d\epsilon}+4x^3 =0.
\end{equation}
To reduce the order of (A8) we introduce $p:=dx/d\epsilon$, which leads 
to the equation
\begin{equation}
p\frac{dp}{dx}-6xp+4x^3 =0.
\end{equation}
Eq. (A9) becomes homogeneous for $z^2 :=p$, since we get
\begin{equation}
\frac{dz}{dx}=\frac{3xz^2 -2x^3}{z^3}.
\end{equation}
Substitution $z:=ux$ into (A10) gives
\begin{equation}
\frac{u^3 du}{-u^4 +3u^2 -2}=\frac{dx}{x}.
\end{equation}
One more substitution $v:=u^2$ turns (A11) into
\begin{equation}
\Bigl(\frac{1}{v-1}-\frac{2}{v-2}\Bigr)dv = \frac{2}{x}dx.
\end{equation}
Solution to (A12) reads
\begin{equation}
\frac{v-1}{(v-1)^2} = Cx^2,
\end{equation}
where $R^1\ni C>0$ is a constant.

\noindent
Making use of of $p=dx/d\epsilon,~p=z^2,~z=ux$ and $v=u^2$ turns (A13) 
into an algebraic equation 
\begin{equation}
\Bigl(\frac{dx}{d\epsilon}\Bigr)^2 - (4x^2 +D)\frac{dx}{d\epsilon} + 4x^4 
+Dx^2 =0,
\end{equation}
where $D:=1/C.$

\noindent
Eq. (A14) splits into two first-order real equations.
One of them has the form (Analysis of the other one can be done by analogy.)
\begin{equation}
2\frac{dx}{d\epsilon}= 4x^2 +D-\sqrt{D(4x^2 +D)} .
\end{equation}
The solution to (A15) reads
\begin{equation}
\epsilon (x) = 2\int\frac{dx}{4x^2 +D-\sqrt{D(4x^2 +D)}}=\frac{1}{A-x-
\sqrt{x^2 +A^2}}+B ,
\end{equation}
where $A=\sqrt{D}/2 $ and $B$ are real constants.

\noindent
Eq. (A16) leads to
\begin{equation}
x(\epsilon)=\frac{A(\epsilon -B)\Bigl(A(\epsilon -B)-1\Bigr) +1}{2 (\epsilon -B)
\Bigl(A(\epsilon -B)-1\Bigr)} .
\end{equation}
Eq. (A17) represents one of the solutions of (A5). It is not defined for 
$\epsilon =B$ because
\begin{equation}
\lim_{\epsilon\rightarrow B-}x(\epsilon) =+\infty,~~~~
\lim_{\epsilon\rightarrow B+}x(\epsilon) =-\infty .
\end{equation}
Since (A17) is not defined for all $\epsilon \in R$ , we conclude 
that the vector field $X_3$ is not complete  on the plane.

One can easily solve the Lie equations corresponding 
to (A1) and (A2). The solutions, respectively, read
\begin{equation}
(t,~x)\longrightarrow (t,~x+b_0)
\end{equation}
and 
\begin{equation}
(t,~x)\longrightarrow (t-rb_1,~x\exp b_1).
\end{equation}
Both (A19) and (A20) describe one-parameter global transformations on $V_p$ 
well defined for any $b_0,b_1 \in R^1$. Therefore, the vector fields $X_1$ 
and $X_2$ are complete on the plane.

\section{REPRESENTATION OF CANONICAL COMMUTATION RELATIONS} 

Let us consider the problem of existence of self-adjoint operators 
$\hat{q}$ and $\hat{p}$ satisfying the relation
\begin{equation}
\hat{q}\hat{p}-\hat{p}\hat{q} = i\hbar\hat{I}~~~~~\text{on}~~~~~
\Omega \subset L^2 [a,b],~~~a,b\in R^1,
\end{equation}
where $L^2 [a,b]$ is the space of square-integrable complex functions 
with the scalar product
\begin{equation}
<\psi_1|\psi_2>= \int_a^b dq\psi^*_1(q)\psi_2(q),
~~~~~~\psi_1,\psi_2 \in L^2[a,b] .
\end{equation}
The dense domain $\Omega$ is defined to be
\begin{equation}
\Omega = \{\varphi\in L^2 [a,b]~|~\varphi\in C^\infty [a,b],~~
\varphi^{(n)}(a)=e^{i \alpha}\varphi^{(n)}(b),~~n=0,1,2,...\},  
\end{equation}
where $0\leq\alpha <2\pi$.

We apply the Schr\"{o}dinger representation to the couple $\hat{q}$ and 
$\hat{p}$:

\noindent
The `position' operator $\hat{q}$  defined as
\begin{equation}
\hat{q}\varphi(q) =q\varphi(q),~~~~~~a\leq q\leq b
\end{equation}
is bounded, so it can be  self-adjoint on $L^2[a,b]$, whereas the `momentum' 
operator $\hat{p}$  ($ \hbar$ is set equal to $1$)
\begin{equation}
\hat{p}\varphi(q)=\frac{1}{i}\frac{d}{dq}\varphi(q),
\end{equation}
is unbounded, so it cannot be defined on the entire $L^2[a,b]$.
It is clear that $\Omega$ is invariant under $\hat{q}, \hat{p}, 
 \hat{q}\hat{p}$ and $\hat{p}\hat{q}$. If $\hat{p}$ is to be a symmetric 
operator on $\Omega$ one should, for example, have
\begin{equation}
<\psi|\hat{p}q\varphi>=<\hat{p}\psi|q\varphi>,~~~~~a\leq q\leq b,
~~~~~\psi,\varphi \in \Omega,
\end{equation}
which by (B2) leads to
\begin{equation}
b\psi^\ast (b)\varphi(b)=a\psi^\ast (a)\varphi(a)~~~\text{for any}~~~a\neq b.
\end{equation}
Eq. (B7) cannot be satisfied, if we only have
\begin{equation}
\psi(a)=\psi(b) e^{i \alpha}~~~\text{and}~~~\varphi(a)=\varphi(b) e^{i \alpha}
~~~\text{for any}~~~a\neq b .
\end{equation}
It can be satisfied in case the functions of $\Omega$ vanish at $a$ and $b$, 
e.g.,
\begin{equation}
\varphi(b)=0=\varphi(a) .
\end{equation}
However, if $\Omega$ includes functions obeying (B9) the deficiency indices 
\cite{RS} of the operator $\hat{p}$  on $\Omega$  are in the relation 
$n_+ \neq n_-$, which means that the operator $\hat{p}$ cannot be  
self-adjoint on such a domain. 

We have not proved that the self-adjoint representation of (B1) does not exist 
for any dense subspace  $\Omega$ of $L^2[a,b]$. We have only shown that 
a self-adjoint representation of (B1) does not exist for the choice 
(B3 - B5) which is used in Sec. III (with $a=0$ and $b=2\pi$).

Let us remind\cite{PU}  the reader  that in case both $\hat{q}$ and $\hat{p}$ 
are bounded operators, the self-adjoint representation of the commutation 
relations (B1) does not exist. In the opposite case, when both $\hat{q}$ and 
$\hat{p}$ are unbounded (and the Weyl relations are satisfied) there exist 
only one (up to unitary equivalence) essentially self-adjoint representation 
of (B1) and it is unitarily equivalent to the Schr\"{o}dinger representation. 

\section{REPRESENTATION OF COMMUTATION RELATIONS OF OBSERVABLES ON HYPERBOLOID 
AND ON PLANE}

\noindent
{\bf The case of hyperboloid:}

Let $L^2[0,2\pi]$ denotes the Hilbert space of square integrable complex 
functions on $[0,2\pi]$ with the inner product
\begin{equation}
<\varphi|\psi>=\int_{0}^{2\pi}d\beta~\varphi^*(\beta)\psi(\beta),
~~~~~~~~\varphi,\psi\in L^2[0,2\pi].
\end{equation} 

In what follows we outline the prove that  representation of $sl(2,R)$ 
algebra defined by ($\hbar$ is set equal to 1)
\begin{equation}
\hat{J}_0\psi(\beta):=\frac{1}{i}\frac{d}{d\beta}\psi(\beta),~~~\beta \in S^1,
~~~\psi\in\Omega ,
\end{equation}
\begin{equation}
{\hat{J}}_1 \psi(\beta):= \Bigl( \cos \beta 
~\hat{J_0} - ({\kappa} -\frac{i}{2})\sin \beta\Bigr)\psi(\beta),
\end{equation}
\begin{equation}
\hat{J}_2 \psi(\beta):= \Bigl( -\sin \beta 
~\hat{J_0} - ({\kappa} -\frac{i}{2})\cos \beta\Bigr)\psi(\beta) ,
\end{equation}
where 
\begin{equation}
\Omega := \{\psi\in L^2[0,2\pi]~|~\psi\in C^\infty[0,2\pi],~\psi^{(n)}(0)
= \psi^{(n)}(2\pi),~ n=0,1,2...\}
\end{equation}
is essentially self-adjoint.

It is clear that $\Omega$ is a dense invariant common domain for 
$\hat{J}_a$ and one can easily verify that each $\hat{J}_a$ is symmetric 
on $\Omega~$ (for $a=0,1,2)$, if $\kappa \in R^1$.  Direct calculations show 
that the domains $D(\hat{J}_a^\ast)$ of the adjoint $\hat{J}_a^\ast$ of 
$\hat{J}_a$ consists of functions $\psi_a$ which  satisfy the condition 
(among others)\cite{DM} 
\begin{equation}
\psi_a(0)=\psi_a(2\pi),~~~~\psi_a \in D(\hat{J}_a^\ast)\subset L^2[0,2\pi]
\end{equation}
for $a=0,1,2$. 

The main idea of the proof is to show\cite{RS} that the only solutions 
to the equations
\begin{equation}
\hat{J^\ast_a}f_{a\pm} = \pm if_{a\pm},~~~~f_{a\pm}\in D(J^\ast_a),~~~~a=0,1,2
\end{equation}
are $f_{a\pm}=0$, i.e. the deficiency indices of $\hat{J_a}$ on $\Omega$  
satisfy $n_{a+}=0=n_{a-}~$ (for $a=0,1,2 $).

The equation (C7) for $a=0$ reads
\begin{equation}
\frac{1}{i}\frac{d}{d\beta}f_{0\pm}(\beta)=\pm if_{0\pm}(\beta)
\end{equation} 
and its general normalized solution is
\begin{equation}
f_{0\pm}(\beta)=C_{0\pm} \exp(\mp\beta),~~~~C_{0+}=\sqrt{2/(1-\exp(-4\pi))},
~~~~C_{0-}=\sqrt{2/(\exp(4\pi)-1)}.
\end{equation}

\noindent
The solutions (C9) does not satisfy (C6). Thus the only solution to (C7) 
is $f_{0\pm}$.

For $a=1$ the equation (C7) can be written as
\begin{equation}
(\cos \beta\frac{d}{d\beta}-r\sin\beta+\lambda_\pm)f_{1\pm}(\beta)=0 ,
\end{equation}
where $r=1/2+\kappa i,~ \kappa\in R,~ \lambda_\pm =1$ or $-1$ for $f_{1+}$ 
or $f_{1-}$, respectively.

\noindent
One can verify that the general solution of (C10) reads
\begin{equation}
f_{1\pm}(\beta)= C_{1\pm}|\cos\beta|^{-r}|\tan(\frac{\beta}{2}+
\frac{\pi}{4})|^{-\lambda\pm} ,
\end{equation}
where $C_{1\pm}$ are complex constants.

\noindent
The immediate calculations show that for $C_{1\pm}\neq 0$
\begin{equation}
\lim \Re f_{1+}(\beta)=\infty =\lim \Im f_{1+}(\beta)~~~\text{as}~~~
\beta\rightarrow \frac{3}{2}\pi\pm
\end{equation}
and
\begin{equation}
\lim \Re f_{1-}(\beta)=\infty =\lim \Im f_{1-}(\beta)~~~\text{as}~~~
\beta\rightarrow \frac{\pi}{2}\pm .
\end{equation}
Therefore $f_{1\pm}$ are not square integrable  and the only 
solutions of (C10) are $f_{1\pm}=0$.

The equation (C7) for $a=2$ has the form
\begin{equation}
(\sin\beta\frac{d}{d\beta}+r\cos\beta -\lambda_\pm)f_{2\pm}(\beta)=0 ,
\end{equation}
where $r$ and $\lambda\pm$ are the same as in (C10).

\noindent
The general solution to (C14) is
\begin{equation}
f_{2\pm}(\beta)=C_{2\pm}|\sin\beta|^{-r}|\tan\frac{\beta}{2}|^{\lambda\pm} ,
\end{equation}
where $C_{2\pm}$ are complex constants.

\noindent
The standard calculations yield 
\begin{equation}
\lim \Re f_{2+}(\beta)=\infty =\lim \Im f_{2+}(\beta)~~~\text{as}~~~
\beta\rightarrow \pi\pm 
\end{equation}
and
\begin{equation}
\lim \Re f_{2-}(\beta)=\infty =\lim \Im f_{2-}(\beta)~~~\text{as}~~~
\beta\rightarrow 0+~~~\text{or}~~~\beta\rightarrow 2\pi -  .
\end{equation}
Thus,  $f_{2\pm}$ are not square integrable unless $C_{2\pm} =0.$ 

This finishes the proof, the detailed verification of consecutive 
steps being left to the reader.

\noindent
{\bf The case of plane:}

Having the proof for the case of the hyperboloid it is very easy to prove that 
the representation of $sl(2,R)$ algebra defined by 
\begin{equation}
\hat{I}_0\phi(\sigma):=\frac{1}{i}\frac{d}{d\sigma}\phi(\sigma),
~~~\sigma \in ]0,2\pi[,~~~\phi\in\Omega_\alpha,~~~\alpha \in [0,2\pi[ ,
\end{equation}
\begin{equation}
{\hat{I}}_1 \phi(\sigma):= \Bigl( \cos \sigma 
~\hat{I_0} - ({\kappa} -\frac{i}{2})\sin \sigma\Bigr)\phi(\sigma),
\end{equation}
\begin{equation}
\hat{I}_2 \phi(\sigma):= \Bigl( -\sin \sigma 
~\hat{I_0} - ({\kappa} -\frac{i}{2})\cos \sigma\Bigr)\phi(\sigma) ,
\end{equation}
where 
\begin{equation}
\Omega_\alpha := \{\phi\in L^2[0,2\pi]~|~\phi\in C^\infty[0,2\pi],
~\phi^{(n)}(0)= e^{i\alpha}\phi^{(n)}(2\pi),~ n=0,1,2...\}
\end{equation}
is essentially self-adjoint too.

Since the functional forms of $\hat{I}_a ~~(a=0,1,2)$ do not depend 
on $\alpha$  and since 
$\exp{(i\alpha)}~\exp{(-i\alpha)}=1$, the operators $\hat{I}_a$ are symmetric 
on $\Omega_\alpha$. An elementary proof includes integration by parts of 
the one side of the equation
\begin{equation}
<\phi_1 | \hat{I_a}\phi_2> = <\hat{I_a}\phi_1 | \phi_2>,~~~~\phi_1, \phi_2 
\in \Omega_\alpha
\end{equation}
followed by application of the property
\begin{equation}
\phi(0) =e^{i\alpha}\phi(2\pi),~~~~\phi \in \Omega_{\alpha}.
\end{equation}
As the functional forms of $\hat{J}_a$ and $\hat{I}_a~~~(a=0,1,2)$ are 
the same, the equations defining the deficiency indices are defined by (C8), 
(C10) and (C14) with general solutions given by (C9), (C11) and (C15), 
respectively. The only difference is that the domains  $D(\hat{I^\ast_a})$ 
of the adjoint $\hat{I^\ast_a}$ of $\hat{I_a}$ are defined by functions 
of $L^2[0,2\pi]$ which must have the property (among others)\cite{DM}
\begin{equation}
<\hat{I_a}\phi | \psi> = <\phi | \hat{I^\ast_a} \psi>,~~~~\forall 
\phi \in \Omega_\alpha,~\psi \in D(\hat{I^\ast_a}), 
\end{equation} 
which leads, after integrating one side of (C24) by parts and making use 
of (C23), to the property
\begin{equation}
\psi(0) =e^{i\alpha}\psi(2\pi),~~~~\forall \psi \in D(\hat{I^\ast_a}).
\end{equation}
In case $a=0,$ Eq. (C9) shows that (C25) cannot be satisfied by $f_{0\pm}$ 
with $C_{0\pm}\neq 0.$

\noindent
In cases $a=1,2$ the solutions $f_{a\pm}$ are singular, so they are not 
in $L^2[0,2\pi]$, unless $C_{a\pm}=0.$

\noindent
Therefore, $n_+(\hat{I_a})=0=n_-(\hat{I_a}) $ for $a=0,1,2,~$ which ends 
the proof.

\section{REPRESENTATION OF COMMUTATION RELATIONS OF REDUCED SET OF 
OBSERVABLES ON PLANE}

We give the proof that the representation of the Lie algebra
\begin{equation}
\{P,K\}=P
\end{equation}
defined by (we set $\hbar =1$)
\begin{equation}
\hat{P}\phi (q):= \frac{1}{i}\frac{d}{dq}\phi(q),~~~~\hat{K}\phi(q) := \Bigl( 
q\frac{1}{i}\frac{d}{dq}+\frac{1}{2i}-\kappa \Bigr)\phi(q),~~~~q\in R^1,
\end{equation}
\begin{equation}
\phi\in \Lambda:=\{\phi\in L^2(R^1)~|~\phi \in C^\infty_0(R^1) \}
\end{equation}
is essentially self-adjoint.

It is easily seen that the representation (D2-D3) is symmetric on common 
invariant dense domain $\Lambda$.

Let us consider the equation
\begin{equation}
\hat{P}^\ast f_\pm(q)=\pm if_\pm(q),~~~~f_\pm \in D(\hat{P^\ast})
\subset L^2(R^1).
\end{equation}
 One can check that the general solution to (D4) reads
\begin{equation}
f_\pm(q)=A_\pm\exp(\mp q) ,
\end{equation}
where $A_\pm$are complex constants.

\noindent
Since $f_\pm$ are not square integrable on $R^1$ the functions $f_\pm$ 
cannot be in $D(\hat{P}^\ast)$ for $A_\pm \neq 0$. Thus the deficiency 
indices of $\hat{P} $ on $\Lambda $ satisfy $n_+(\hat{P})=0=n_-(\hat{P})$, 
which means\cite{RS} that $\hat{P}$ is essentially self-adjoint on $\Lambda$.

In case of $\hat{K}$ operator the equations for finding the deficiency indices 
$n_+(\hat{K})$ and $n_-(\hat{K})$ read
\begin{equation}
\hat{K}^\ast g_\pm(q) = \pm ig_\pm(q),~~~~g_\pm \in D(\hat{K}^\ast)
\subset L^2(R^1)
\end{equation}
which can be rewritten as
\begin{equation}
q\frac{d}{dq}g_\pm(q) + \omega_{\pm} g_\pm(q) =0,~~~~
\omega_{\pm} :=\frac{1}{2}\pm 1 -\kappa i .
\end{equation}
The general solution to (D7) has the form
\begin{equation}
g_\pm (q)= B_\pm ~q^{-\omega_{\pm}} , 
\end{equation}
where $B_\pm$ are complex constants.

\noindent
One can verify that $g_\pm$ are in $L^2(R)$ only for $B_\pm =0$,  
which means that $n_+(\hat{K})=0=n_-(\hat{K})$.
Therefore, Eqs. (D2) and (D3) define an essentially self-adjoint 
representation of (D1) algebra.


\begin{references}

\bibitem{GW} G. Jorjadze  and W. Piechocki,  
             Phys. Lett. B461 ( 1999 ) 183; Theor. Math. Phys. 
             118 ( 1999 ) 183.
           
\bibitem{ES}E. Schr\"{o}dinger, Expanding Universes ( Cambridge University 
             Press, Cambridge, 1956 ).

\bibitem{HE} S. W. Hawking and G. F. R. Ellis, The Large Scale Structure of 
             Space-Time ( Cambridge University Press, Cambridge, 1973 ). 
               
 
 \bibitem{VA} V. I. Arnold, Mathematical Methods for Classical Mechanics 
             ( Springer - Verlag, New York, Inc., 1978 ).
 
 \bibitem{CJI} A. A. Kirillov, Elements of the Theory of Representations 
               ( Springer, Berlin, 1976 ).
             
 \bibitem{LP} L. Pontrjagin, Topological Groups ( Princeton University Press, 
              Princeton, 1946 ).
 
\bibitem{BR} A. O. Barut and R. R\c{a}czka, Theory of Group Representations 
             and Applications ( World Scientific Publishing, Singapore, 1986 ).
              

\bibitem{H4} G. Jorjadze and W. Piechocki, Phys. Lett. B476 ( 2000 ) 134.


\bibitem{JS} J. M. M. Senovilla, Gen. Relativ. Gravit. 30 ( 1998 ) 701. 


\bibitem{NHI} N. H. Ibragimov, Elementary Lie Group Analysis and Ordinary 
              Differential Equations ( John Wiley and Sons, Chichester, 1999 ).
              

\bibitem{PU} C. R. Putnam, Commutation Properties of Hilbert Space Operators 
             and Related Topics ( Springer-Verlag, Berlin, 1967 ).
             
\bibitem{DM} L. Debnath and P. Mikusi\'{n}ski, Introduction to Hilbert Spaces 
             with Applications ( Academic Press, San Diego, 1999 ).


\bibitem{RS} M. Reed and B. Simon, Methods of Modern Mathematical Physics 
            ( Academic Press, New York, 1975 ).


\end{references}
\end{document}